\magnification=1200
\centerline{A GENERALIZED DETAILED BALANCE RELATION.}
\bigskip
\centerline{by David Ruelle\footnote{$\dagger$}{Math. Dept., Rutgers University, and 
IHES, 91440 Bures sur Yvette, France. email: ruelle@ihes.fr}.}
\bigskip\bigskip\bigskip\bigskip\noindent
	{\leftskip=1.8cm\rightskip=2cm{\sl Abstract:}
\medskip\noindent
Given a system $M$ in a thermal bath we obtain a generalized detailed balance relation for the ratio $r=\pi_\tau(K\to J)/\pi_\tau(J\to K)$ of the transition probabilities $M:J\to K$ and $M:K\to J$ in time $\tau$.  We assume an active bath, containing solute molecules in metastable states.  These molecules may react with $M$ and the transition $J\to K$ occurs through different channels $\alpha$ involving different reactions with the bath.  We find that $r=\sum p^\alpha r^\alpha$, where $p^\alpha$ is the probability that channel $\alpha$ occurs, and $r^\alpha$ depends on the amount of heat (more precisely enthalpy) released to the bath in channel $\alpha$.\par}
\vfill\eject
	The theory of dynamical systems has recently been used in certain questions of biology [1],[4],[6].  The present paper originates with the author's desire to clarify the application of detailed balance relations, specifically as used in [4] for a large system $M$.  We begin with the basic ideas of statistical mechanics, and use deterministic dynamics, which is more fundamental than the stochastic methods often used to study nonequilibrium [3],[2].  We feel that we can thus make more precise the scope of the results obtained and the limits to their validity.  The system $M$ in which we are interested is far from equilibrium, but the specific detailed balance problem which we discuss reduces to a problem of equilibrium statistical mechanics, because of the use of metastable states and time reversal symmetry.  This is why a stochastic dynamical model is not needed here.  Our arguments are thus not opposed to the ideas of Jarzynski [9] which actually uses Hamiltonian dynamics, and Crooks [2] (and the earlier fluctuation formula [5], and theorem [7], see also [10],[12],[13],[15]) but indicate how an application of detailed balance can be made by invoking more directly the basic laws of classical Hamiltonian mechanics and equilibrium statistical mechanics.  Mixing properties of the physical dynamics are used implicitly, but we avoid stronger Markovian stochastic assumptions.  [The the chaotic hypothesis formulation of Gallavotti and Cohen [7] has the virtue of using determinist dynamics and leading to rigorous results, but it remains usually an uncontrolled approximation].
\medskip
	The main result of this paper is (1.6) below, with its corollary (1.7).  Our discussion will remain at the level of theoretical rather than mathematical physics.  We shall in particular avoid discussing the limits implied in the use of metastable states.  We feel that forcing the problem at hand in the straitjacket of some limited mathematical formalism would be counterproductive at this time.
\medskip
	The present paper follows an earlier related note [16] where we did not explicitly take into account the chemical reactions of $M$ with the bath.  That note has benefitted from useful comments by Pierre Gaspard [8] to which we refer for a clear and concise introduction to the general area of ``biology and nonequilibrium''.
\medskip
	We start our paper with a physical discussion bypassing dynamical considerations (Section 1), make a number of remarks (Section 2), and then analyze detailed balance from the point of view of deterministic dynamics (Section 3).
\bigskip\noindent
{\bf Acknowledgments.}
\medskip
	I thank the referees who pointed out to an incorrect derivation in an earlier version of the present article, and suggested a number of improvements.  Their comments have led to a complete rewriting of Section 3 of the paper.
\vfill\eject
\noindent
{\bf 1. Physical discussion.}
\medskip
	The following situation is suggested by biology [4]: a system $M$ (large molecule or perhaps bacterial cell) is immersed in a large bath (water with solutes which are small molecules).  Temperature, pressure, and chemical potentials of the solutes are kept constant.  The bath is contained in a bounded region $\Lambda\subset{\bf R}^3$.  We choose a region $V$ around $M$, with $V$ somewhere in the middle of $\Lambda$.  We think of $V$ as small and $\Lambda$ as large (we may let $\Lambda\nearrow{\bf R}^3$).  We describe $M+\,$bath by a Hamiltonian $H_\Lambda$ which may fix or constrain the position of $M$ in $V$.  We assume that $H_\Lambda$ is invariant under a time-reversal operation $T$.
\medskip
	Let $\pi_\tau(J\to K)$ be the conditional probability that, starting from a state $J$, the system $M$ evolves in time $\tau$ to a state $K$.  We shall obtain a detailed balance relation (see $(1.6)$ below) generalizing the standard equality
$$	{\pi_\tau(K\to J)\over\pi_\tau(J\to K)}=\exp[\beta(G(K)-G(J))]\eqno{(1.1)}   $$
where $\beta$ is the inverse temperature and $G$ denotes the Gibbs free energy.  The proof follows as usual from the assumed time reversal invariance of the microscopic dynamics.  Our discussion and proposed answer will remain formal in this Section.  Section 3 will be based on deterministic dynamics for the system $M+\,$bath.  For the rest the ideas we shall use are rather standard.
\medskip
	Let us now specify the chosen setup, and in particular the kind of state of the system $M$ which we consider.  A state $J$ of $M$ should describe the inside of a small region $V$ around $M$ by the rules of equilibrium statistical mechanics, with an important qualification: $J$ should be a metastable state with long lifetime compared with local equilibration times (thermal and pressure equilibration in the bath).  The Gibbs free energy $G(J)$ should make sense.  We shall give in Section 3 our description of metastable states.
\medskip
	We assume now that we have an active bath, so that the transition $J\to K$ can proceed through a number of channels $\alpha$.  Each channel corresponds to the interaction of $M$ with solute molecules (small molecules that may be in metastable states) with a set $\alpha^{\rm in}$ of ingoing molecules being replaced by a set $\alpha^{\rm out}$ of outgoing molecules.  We let $\Delta^\alpha G=G(\alpha^{\rm out})-G(\alpha^{\rm in})$ be the change of free energy of solute molecules associated with the channel $\alpha$.  Clearly, the relation $(1.1)$ will have to be replaced by a generalized detailed balance relation taking into account not only $\Delta G=G(K)-G(J)$ but also the $\Delta^\alpha G$.
\medskip
	We are using free energies $G(\alpha^{\rm in,out})$ away from $M$ but we assume that the change $(J,\alpha^{\rm in})\to(K,\alpha^{\rm out})$ takes place at the position of $M$.  This fits with the notion that the lifetimes of $J,K$ (and therefore $\tau$) are large with respect to local equilibration times.  One of the channels, say $\alpha=0$, may correspond to $\alpha^{\rm in}$ and $\alpha^{\rm out}$ being empty.
\medskip
	We write the transition probability $J\to K$ as a sum over channels:
$$	\pi_\tau(J\to K)=\sum_\alpha\pi_\tau^\alpha(J\to K)\eqno{(1.2)}   $$
For the channel $\alpha$ the relation $(1.1)$ takes the form
$$	{\pi_\tau^{-\alpha}(K\to J)\over\pi_\tau^\alpha(J\to K)}
	=\exp[\beta(\Delta G+\Delta^\alpha G)]\eqno{(1.3)}   $$
Here $-\alpha$ is obtained from $\alpha$ by interchanging $\alpha^{\rm in}$ and $\alpha^{\rm out}$.  Note that $\Delta^\alpha G$ takes into account the imposed concentrations of the $\alpha^{\rm in}$ and $\alpha^{\rm out}$ solutes via their chemical potentials.  [The chemical potential $\mu_A$ for a solute $A$ is such that $\exp[\beta\mu_A]$ is, for a dilute solution, proportional to the concentration $m_A$, i.e., $\exp[\beta\mu_A]\approx\exp[\beta\mu_A^0].m_A$.  We have $\Delta^\alpha\mu=0$ for the channel $\alpha=0$, and $\exp[\beta\Delta^\alpha\mu]=0$ if some solute in $\alpha^{\rm out}$ has concentration 0].
\medskip
	Introduce the probabilities
$$	p^\alpha=\pi_\tau^\alpha(J\to K)/\pi_\tau(J\to K)\eqno{(1.4)}   $$
[these probabilities depend on $\tau$, temperature, pressure, and chemical potentials of the solutes].  From $(1.2),(1.3),(1.4)$ we obtain
$$	\pi_\tau(K\to J)=\sum_\alpha\pi_\tau^{-\alpha}(K\to J)
	=\sum_\alpha\pi_\tau^\alpha(J\to K).\exp[\beta(\Delta G+\Delta^\alpha G)]   $$
$$=\pi_\tau(J\to K)\sum_\alpha p^\alpha\exp[\beta(\Delta G+\Delta^\alpha G)]\eqno{(1.5)}$$
We may write $\beta\Delta G=-\Delta S+\beta\Delta H$ where $\Delta S$ is the change of entropy of $M$ and $\Delta H$ is the change of enthalpy of $M$ in the transition $J\to K$.  [Enthalpy is energy + a $pressure$ $\times$ $volume$ term which can usually be ignored in the nearly incompressible situations discussed here].  Similarly, we write $\beta\Delta^\alpha G=-\Delta^\alpha S+\beta\Delta^\alpha H$.  With this notation we have
$$	{\pi_\tau(K\to J)\over\pi_\tau(J\to K)}=\exp[-\Delta S+\beta\Delta H]
	\sum_\alpha p^\alpha\exp[-\Delta^\alpha S+\beta\Delta^\alpha H]\eqno{(1.6)}   $$
Note that $-\Delta H$ is the contribution of the system $M$ to the change of enthalpy of the bath in the transition $J\to K$, while $-\Delta^\alpha H$ is the contribution of the solutes transition $\alpha^{\rm in}\to\alpha^{\rm out}$ corresponding to the channel $\alpha$.
\medskip
	Using the convexity of $\exp$ we obtain from $(1.6)$ the inequality
$$	{\pi_\tau(K\to J)\over\pi_\tau(J\to K)}
	\ge\exp[-\Delta S+\beta\Delta H+\sum_\alpha p^\alpha(-\Delta^\alpha S+\beta\Delta^\alpha H)]   $$
$$	=\exp[-\langle\Delta S\rangle-\beta\langle\Delta Q\rangle]\eqno{(1.7)}   $$
where $\langle\Delta S\rangle=\Delta S+\sum p^\alpha\Delta^\alpha S$, $\langle\Delta Q\rangle=-\Delta H-\sum p^\alpha\Delta^\alpha H$, so that $\langle\Delta Q\rangle$ is the average amount of heat (more precisely enthalpy) transferred to the bath in the transition $J\to K$.
\vfill\eject
\noindent
{\bf 2. Remarks.}
\medskip
	(a) Our relation $(1.7)$ is close to a relation (see [4], equation (8)) used by England in his discussion of self-replication, but it is more explicit, and $(1.6)$ is more precise.  One can also relate $(1.6)$ and $(1.7)$ to the discussion in [4].  Note that while the discussion in [4] leads to a maximum dissipation of free energy in the environment, the discussion in [6] leads to minimum dissipation.  There is no contradiction because the problems considered in [4] and [6] are different.
\medskip
	(b) The generalized detailed balance relation $(1.6)$ differs from $(1.1)$ and also $(1.7)$ by the fact that that it contains not only thermodynamic quantities ($\Delta S,\Delta^\alpha S,\Delta H,\Delta^\alpha H$), but also the probabilities $p^\alpha$ which are ratios of reaction rates far from equilibrium.  In spite of this the relations $(1.6)$ are useful.
\medskip
	(c) Using the equality $(1.6)$ and an estimate of the dominant $p^\alpha$ leads to an estimate of $\pi_\tau(K\to J)/\pi_\tau(J\to K)$, not just a lower bound as provided by $(1.7)$.  Note also that, if one of the $\alpha^{\rm out}$ solutes has zero concentration (so that $\Delta^\alpha\mu=-\infty$) the inequality $(1.7)$ becomes $\pi_\tau(K\to J)/\pi_\tau(J\to K)\ge0$ which is trivial, while $(1.6)$ may remain useful.
\medskip
	(d) Instead of the probabilities $p^\alpha$ for $(J,\alpha^{\rm in})\to(K,\alpha^{\rm out})$ consider now the probabilities $\bar p^\alpha$ corresponding to the reverse transitions $(K,\alpha^{\rm out})\to(J,\alpha^{\rm in})$.  We have
$$	\bar p^\alpha=\pi^{-\alpha}(K\to J)/\pi(K\to J)   $$
so that, using $(1.3)$ and $(1.5)$,
$$	{\bar p^\alpha\over p^\alpha}
	={\pi_\tau^{-\alpha}(K\to J)\over\pi_\tau^\alpha(J\to K)}\Big/{\pi_\tau(K\to J)\over\pi_\tau(J\to K)}
={\exp[\beta(\Delta G+\Delta^\alpha G)]\over\sum_\gamma p^\gamma\exp[\beta(\Delta G+\Delta^\gamma G)]}   $$
or
$$\bar p^\alpha={p^\alpha\exp[\beta\Delta^\alpha G]\over\sum_\gamma p^\gamma\exp[\beta\Delta^\gamma G]}   $$
i.e., the probabilities $\bar p^\alpha$ are obtained by normalizing the weights $p^\alpha\exp[\beta\Delta^\alpha G]$.  In particular, the transition $J\to K$ may be dominated by some channels $\alpha$, while the reverse transition $K\to J$ is dominated by different channels.
\medskip
	(e) From $(1.7)$ we have
$$	1\ge\exp[-\langle\Delta S\rangle_{J\to K}-\langle\Delta S\rangle_{K\to J}
	-\beta\langle\Delta Q\rangle_{J\to K}-\beta\langle\Delta Q\rangle_{K\to J}]   $$
so that
$$	\langle\Delta S\rangle_{J\to K}+\langle\Delta S\rangle_{K\to J}
	+\beta\langle\Delta Q\rangle_{J\to K}+\beta\langle\Delta Q\rangle_{K\to J}\ge0   $$
i.e., in the average, free energy of the solutes is transferred to enthalpy of the bath.
\medskip
	(f) If only the channel $\alpha=0$ contributes to $(1.6)$ we recover the relation (1.1).
\medskip
	The above remarks suggest that a useful application of detailed balance to biological problems requires some knowledge or guess about the relevant $p^\alpha$.  We shall not further discuss biological applications of detailed balance in this paper.
\bigskip\noindent
{\bf 3. Towards a proof of (generalized) detailed balance.}\footnote{*}{This Section has been rewritten, making the conceptual structure more explicit and removing an incorrect derivation of (1.5), (1.6) in an earlier version.}
\medskip
	Detailed balance, like Landauer's principle [11], involves a system outside of equilibrium interacting with a ``bath'' described by equilibrium statistical mechanics at a certain nonzero temperature.  A rigorous analysis of such situations is difficult, it should involve quantum mechanics, and we shall not face it here.  As already said the present paper discusses detailed balance from the point of view of classical mechanics and statistical mechanics, avoiding Markovian stochastic assumptions.  Our ambition in this Section is to make physical sense of the problem and to locate the points (essentially involving equilibrium statistical mechanics) where progress is needed to obtain a rigorous understanding.
\medskip
	Detailed balance relations follow from a combination of the principles of equilibrium statistical mechanics and the basic laws of Hamiltonian mechanics (particularly time reversal invariance).  Markovian stochasticity may or may not be a good approximation to the Hamiltonian dynamics, and it is thus desirable to realize that it is not a necessary assumption.  (It is physically reasonable for particle diffusion processes in the bath).
\medskip
	The system of interest here (system $M+\,$bath) is enclosed in a region $\Lambda\subset{\bf R}^3$.  It is composed of particles (atoms or molecules, this will be discussed later) described by classical mechanics with a Hamiltonian $H(P,Q)$. The Hamiltonian is a function on the phase space $\Phi_\Lambda$: the points of $\Phi_\Lambda$ have the form $\Omega=(P,Q)$ where $P$ denotes momentum variables and $Q$ position variables.  The Hamiltonian has the usual form: sum of kinetic energies and interparticle potentials.  We assume that the interparticle potentials have the properties (moderate range, small distance repulsion) required for equilibrium statistical mechanics to hold.  This is an essential assumption, which leads to the thermodynamic behavior of macroscopic systems.
\medskip
	Equilibrium statistical mechanics defines equilibrium states $\rho$, which are probability measures on the phase space $\Phi_\Lambda$, proposed to describe probabilistic situations called ``equilibrium'' (at given temperature, pressure, and chemical composition) for a system of particles in $\Lambda$.
\medskip
	There are several ways to define the equilibrium measure $\rho$, corresponding to different {\it ensembles}.  These definitions are expected to be equivalent for large systems (equivalence of ensembles, see [14]).  In the microcanonical ensemble, the equilibrium measure is the normalized volume at fixed energy in the phase space $\Phi_\Lambda$.  In the canonical ensemble the temperature $\beta^{-1}$ is fixed, and the equilibrium state $\rho=\rho_\Lambda$ is given by
$$	\rho_\Lambda(P,Q)\,dP\,dQ=Z^{-1}\exp[-\beta H(P,Q)]\,dP\,dQ\quad,\quad
	Z=\int\exp[-\beta H(P,Q)]\,dP\,dQ   $$
If $R\subset\Phi_\Lambda$ is a region of phase space one may write its weighted phase space volume as
$$	\int_R\exp[-\beta H(P,Q)]\,dP\,dQ=\exp[-\beta F(R)]   $$
where $F(R)$ is called the Helmholtz free energy of $R$.  Therefore $\rho_\Lambda(R)$ is proportional to $\exp[-\beta F(R)]$ at given temperature $\beta^{-1}$ and 3-volume $|\Lambda|$.
\medskip
	In the isothermal-isobaric ensemble the temperature $\beta^{-1}$ and the pressure $p$ are fixed, the volume $|\Lambda|$ is allowed to fluctuate and an integration $\int e^{\beta p|\Lambda|}\,d|\Lambda|$ is introduced so that $Z$ is replaced by $\int Z e^{\beta p|\Lambda|}\,d|\Lambda|$.  The Helmholtz free energy is replaced by the Gibbs free energy $G=F+p|\Lambda|$ [$|\Lambda|$ is here a mean value, for large systems the 3-volume of $\Lambda$ fluctuates relatively little in the isothermal-isobaric ensemble].  When the temperature and the pressure are fixed, a natural definition of weighted phase space volume of a region $R$ is thus $\exp[-\beta G(R)]$ where $G(R)$ is the Gibbs free energy of $R$, and the equilibrium probability $\rho(R)$ is proportional to $\exp[-\beta G(R)]$.  If we change $R$ at fixed temperature and pressure, the relative change in weighted phase space volume is thus
$$	\exp[\beta(G(R)-G(R'))]=\rho(R')/\rho(R)   $$
In view of the equivalence of ensembles the same ratio $\rho(R')/\rho(R)$ at given $\beta$ and $p$ can also be computed from the canonical ensemble at given $\beta$ and $\Lambda$ provided $|\Lambda|$ corresponds to the mean value from the iso-thermal-isobaric ensemble at the given value of $p$.  Note however that we shall be using equivalence of ensembles beyond what has been rigorously proved at this time.
\medskip
	We turn to the relation between equilibrium statistical mechanics and the time evolution of our system of particles in $\Lambda$.  This time evolution $(f^t)$ is given by the Hamiltonian evolution equations (complemented by reflection of the particles at the boundary of $\Lambda$).  Note that $(f^t)$ preserves the phase space volume $dP\,dQ$ and the total energy, and therefore also the equilibrium measure $\rho$, whether it is given by the microcanonical or the canonical ensemble.  The time reversal $T$ acting on $\Phi_\Lambda$ by $T(P,Q)=(-P,Q)$ satisfies $Tf^t=f^{-t}T$, and also preserves $\rho$.  The fact that phase space volume is preserved by time evolution is reflected in the preservation of the entropy (essentially the logarithm of the phase space volume) and the preservation of the Helmholtz and Gibbs free energies (linear combinations of entropy, energy, and $pressure\times volume$ $|\Lambda|$).
\medskip
	The application of detailed balance which we want to discuss involves molecules, and will force us to change somewhat our presentation of equilibrium statistical mechanics.  A molecule is a probabilistically described collection of atoms which remain spatially related over a fairly long time.  The interaction between the atoms of a molecule with each other and with other molecules can destroy the relations that define the molecules.  (For example the structure of a molecule can be rearranged: one metastable state goes over to another metastable state).  Experimentally molecules can be very long-lived.  In certain situations we can usefully apply equilibrium statistical mechanics to long-lived molecules considered as interacting particles.  One can thus describe a liquid with given concentrations of solute molecules (although these concentrations are, strictly speaking, not constant and would change noticeably over sufficiently long times).  In particular one can attribute chemical potentials to various solutes although this concept makes sense only within a certain approximate description of the solute molecules.  The situation considered lacks a proper mathematical study (and a quantum description would really be needed here).  But the physics is sufficiently clear to allow us to proceed.  We shall call $\tilde\rho$ the equilibrium probability measure for $M+\,$bath (and solutes) as indicated above.  The measure $\tilde\rho$ is thus concentrated on only part of the full phase space $\Phi_\Lambda$.  On this part $\tilde\rho$ differs from $\rho$ because the concentration of the solute molecules is imposed, corresponding to imposed values for chemical potentials.  We assume that $\tilde\rho$ is invariant under the time reversal $T$.
\medskip
	In the case of interest here, equilibrium probabilities are attributed by $\tilde\rho$ to all kinds of configurations of the system $M$ in $V$.  We think of the system $M$ as a larger structure than the solute molecules.  Different states $J,K,\dots$ of $M$ will correspond to different subsets $R_J,R_K,\dots$ of the phase space corresponding to the inside of $V$, see below.  Outside of the system $M$, $\tilde\rho$ describes water and solutes.
\medskip
	We have noted that the Hamiltonian time evolution $(f^t)$ preserves various global quantities: entropy, energy, free energies.  Because of our assumptions on interparticle potentials (which permit a localized definition of energy) and our description of a system of molecules by statistical mechanics we can attribute entropy, energy, and free energies to subsystems occupying local subregions of ${\bf R}^3$.  In particular for a large bath with given temperature, pressure, and chemical composition, one can make sense of the Gibbs free energy of a given metastable solute molecule (in the limit of a large lifetime of the molecule).  The Gibbs free energy of $M$ in a metastable state $J,K,\dots$ is also defined (we allow the states $J,K,\dots$ to contain different numbers of atoms).
\medskip
	For the system $M+\,$bath as described above we shall allow chemical reactions only in contact with $M$.  [No chemical reaction takes place within the bath: experimentally this is equivalent to keeping the chemical composition of the bath constant].
\medskip
	What we have said shows that to discuss detailed balance we have to go beyond the use of an equilibrium state $\rho$ on a phase space $\Phi_\Lambda$ describing atoms.  Under suitable conditions (temperature, pressure and chemical composition) we use an ``equilibrium measure'' $\tilde\rho$ for long-lived molecules instead of $\rho$.  The probability measure $\tilde\rho$ is carried by only a small subset of the phase space $\Phi_\Lambda$, and the chemical potentials of solutes which we must use make sense only on this small subset.  But detailed balance involves only ratios of probabilities, which can be related between $\tilde\rho$ and $\rho$.  We can also use $\tilde\rho$ and $\rho$ to estimate differences in free energies like $\Delta G,\Delta^\alpha G$.
\medskip
	As stated above, equilibrium statistical mechanics considers particle systems with interparticle potentials satisfying suitable conditions (repelling at short distances, suitably decaying at large distances) and proves thermodynamic behavior for large systems.  For the microcanonical ensemble this goes as follows.  Let the 3-volume of the region $\Lambda$ enclosing the system $M+\,$bath be proportional to the number $n$ of particles it contains (we assume only one kind of particles) and let the total energy $E$ also be proportional to $n$.  Then the energy shell $\Sigma_E\subset\Phi_\Lambda$ has a phase space volume $|\Sigma_E|$ which grows exponentially with $n$.  In other words the entropy $\log|\Sigma_E|$ grows proportionally to $n$, i.e., the entropy is an extensive variable.  (By definition the number of particles $n$, the total energy $E$, and the 3-volume of $\Lambda$ are extensive).  The total Gibbs free energy $G$ is also extensive.  The inverse temperature $\beta$ and the pressure $p$ are intensive variables: they tend to a limit for large systems.  These notions extend to several kinds of particles in a straightforward manner.  The standard thermodynamic facts just recalled are theorems depending essentially on suitable assumptions for interparticle potentials.
\medskip
	We can now return to the relation (1.5) or (1.6) and make sense of its proof in terms of classical statistical mechanics and Hamiltonian dynamics.  Since (1.2) and (1.4) are just definitions we need to understand (1.3) which is basically the standard detailed balance relation (1.1).
\medskip
	We describe the states $J,K$ of $M$ by sets $R_J,R_K\subset\Phi_\Lambda$ with characteristic functions $\chi_J,\chi_K:\Phi_\Lambda\to\{0,1\}$.  The state $J$ corresponds to the probability measure $\chi_J.\tilde\rho/\tilde\rho(\chi_J)$ on $\Phi_\Lambda$, i.e., the equilibrium state $\tilde\rho$ is restricted to $R_J$ and then normalized.  Similarly for the state $K$.  To be able to interpret $J$ and $K$ as long-lived metastable states of $M$ we make the following assumptions:
\medskip\noindent
(a)  $R_J$ and $R_K$ are localized around the system $M$ (i.e., $\chi_J(\Omega)=1,\chi_K(\Omega)=1$ impose conditions only on the phase space coordinates of particles in or around $V$).
\medskip\noindent
(b)  $R_J$ and $R_K$ are invariant under the time reversal $T$.
\medskip\noindent
(c)  $R_J$ and $R_K$ are almost invariant under $(f^t)$ in the sense that the periods of time such that $f^t\Omega$ stays in $R_J$ or $R_K$ (or $\Phi_\Lambda\backslash R_J$ or $\Phi_\Lambda\backslash R_K$) are large compared with local equilibration times (to be specified below).
\medskip
	We first discuss (1.3) when there are no chemical interactions of $M$ with solute molecules in the bath.  (Here we could thus use $\rho$ instead of $\tilde\rho$).  With the probabilistic description of the system $M+\,$bath given by $\tilde\rho$, we write the probability that $M$ is in the state $J$ as $\tilde\rho(\chi_J)$.  The conditional probability that $M$, starting in state $J$ at time 0, has moved to state $K$ at time $\tau$ is
$$	\pi_\tau(J\to K)=\tilde\rho(\chi_J.(\chi_K\circ f^\tau))/\tilde\rho(\chi_J)\eqno{(3.1)}   $$
The conditional probability of the reverse transition $K\to J$ is
$$	\pi_\tau(K\to J)=\tilde\rho(\chi_K.(\chi_J\circ f^\tau))/\tilde\rho(\chi_K)   $$
The assumed invariance under time reversal $T$, and $(f^t)$ invariance, gives
$$	\tilde\rho(\chi_K.(\chi_J\circ f^\tau))=\tilde\rho(\chi_K.(\chi_J\circ f^{-\tau}))
	=\tilde\rho((\chi_K\circ f^\tau).\chi_J)=\tilde\rho(\chi_J.(\chi_K\circ f^\tau))\eqno{(3.2)}   $$
so that
$$	{\pi_\tau(K\to J)\over\pi_\tau(J\to K))}={\tilde\rho(\chi_J)\over\tilde\rho(\chi_K)}
	={\rho(\chi_J)\over\rho(\chi_K)}\eqno{(3.3)}   $$
where the last equality reflects the proportionality of $\tilde\rho$ and $\rho$ on the states considered.  Therefore (3.3) is also the ratio of volumes in phase space associated with $J$ and $K$.  The long lifetime of $J$ and $K$ means that these states of $M$ are in equilibrium with their surroundings hence have equal values for the intensive variables: temperature and pressure.  The corresponding phase space volumes in $\Phi_\Lambda$ are proportional to $\exp[-\beta G(J)]$ and $\exp[-\beta G(J)]$, being expressed in terms of the Gibbs free energies of $J,K$ at the given temperature and pressure.  These facts are in physical agreement with thermodynamics and yield the standard form
$$	{\pi_\tau(K\to J)\over\pi_\tau(J\to K))}=\exp[\beta(G(K)-G(J)]=\exp[\beta\Delta G]   $$
of detailed balance.  It must however be strongly pointed out that a mathematical study of the equilibrium statistical mechanics of metastable states like $J,K$ as used above is currently missing.  Such a study should provide a rigorous link between phase space volumes associated with metastable states, and Gibbs free energies.
\medskip
	We now want to take into account the interaction of $M$ with solute molecules to justify the general form of (1.3):
$$	{\pi_\tau^{-\alpha}(K\to J)\over\pi_\tau^\alpha(J\to K)}=\exp[\beta(\Delta G+\Delta^\alpha G)]\eqno{(1.3)}   $$
We leave out a discussion of metastability for solute molecules, which should go along the same lines as the discussion of the metastability of $J$ and $K$ above.  The term $\pi_\tau^\alpha(J\to K)$ in (1.3) is the $\tilde\rho$-probability for $\Omega\in\Phi_\Lambda$, conditioned on $\chi_J(\Omega)=1$, that some family $\alpha^{\rm in}$ of solute molecules in the bath will be absorbed by $M$ while a set $\alpha^{\rm out}$ of molecules are emitted so that $\chi_K(f^\tau\Omega)=1$.  Remember that $f^\tau$ preserve $\rho$, but that the states $(J,\alpha^{\rm in}),(K,\alpha^{\rm out})$ are equilibrium states for $\tilde\rho$, not for $\rho$.
\medskip
	For a given phase space location of $\alpha^{\rm in}$ we can write $\pi_\tau^\alpha(J\to K)$ like (3.1) as a quotient $N^\alpha/D^\alpha$.  The denominator now is
$$	D^\alpha=\tilde\rho(\chi_J)\times\tilde\rho(\alpha^{\rm in})   $$
where the factor $\tilde\rho(\chi_J)$ is because of the conditioning on $\chi_J(\Omega)=1$, and $\tilde\rho(\alpha^{\rm in})\sim\exp[-\beta G(\alpha^{\rm in})]$ takes into account the $\rho$-volume of individual solute molecules and their their concentrations.
\medskip
	Writing $\pi_\tau^{-\alpha}(K\to J)=\bar N^\alpha/\bar D^\alpha$, we see that the numerators $\bar N^\alpha$ and $N^\alpha$ are equal because of $T$ and $(f^t)$ invariance as in (3.2), and the fact that $\bar N^\alpha$ and $N^\alpha$ have the same chemical potential factors.  Therefore the ratio
$$	{D^\alpha\over\bar D^\alpha}
	={\tilde\rho(\chi_J)\times\tilde\rho(\alpha^{\rm in})\over\tilde\rho(\chi_K)\times\tilde\rho(\alpha^{\rm out})}
	=\exp[\beta(\Delta G+\Delta^\alpha G)]   $$
is independent of the phase space location of $\alpha^{\rm in}$ and $\alpha^{\rm out}$.  From this (1.3) follows.
\medskip
	Note that (3.1) expresses that fluctuations of the phase space point $\Omega$ in the state $J$ bring this phase space point after time $\tau$ within the range of $K$.  Since the energy and volume of $M$ are not the same in states $J$ and $K$, there are exchanges in the bath to equalize temperature and pressure.  We have assumed that the times for these local equilibrations are short with respect to the time $\tau$ involved in the transition $J\to K$.  This assumption makes sense for deterministic dynamics and does not require Markovian stochasticity, which may not be a good approximation for $J\to K$.
\medskip
	When reaction with solute molecules is involved in the transition $J\to K$, we assume that these molecules $\alpha^{\rm in},\alpha^{\rm out}$ are mostly in a neighborhood $W$ of the system $M$, such that local equilibration of temperature and pressure takes place over the region $W$ in times small with respect to $\tau$.
\medskip
	Detailed balance involves metastable states at given temperature and pressure.  Therefore we must require as above that local equilibration times be short with respect to the lifetime of metastable states.  For our purposes these requirements also appear physically to be sufficient.
\vfill\eject
\noindent
{\bf References.}
\medskip\noindent
[1] D. Andrieux and P. Gaspard.  ``Nonequilibrium generation of information in copolymerization processes.''  Proc. Natl. Acad. Sci. USA. {\bf 105},9516-9521(2008).
\medskip\noindent
[2] G.E. Crooks.  ``Entropy fluctuation theorem and the non equilibrium work relation for free energy differences.''  Phys. Rev. E {\bf 60},2721-2726(1999).
\medskip\noindent
[3] S.R. de Groot and P. Mazur.  {\it Nonequilibrium thermodynamics.}  Dover, New York, 1984.
\medskip\noindent
[4] J.L. England.  ``Statistical physics of self-replication.''  J. Chem. Phys. {\bf 139},121923 (2013).
\medskip\noindent
[5] D.J. Evans and D.J. Searles.  ``Equilibrium micro states which generate second law violating steady states.''  Phys. Rev. E{\bf 50},1645-1648(1994).
\medskip\noindent
[6] R.F. Fox.  ``Contributions to the theory of thermostatted systems II: Least dissipation of Helmholtz free energy in nano-biology.''  arXiv:1503.03350.
\medskip\noindent
[7] G. Gallavotti and E.G.D. Cohen.  "Dynamical ensembles in nonequilibrium
statistical mechanics."  Phys. Rev. Letters {\bf 74},2694-2697(1995).
\medskip\noindent
[8] P. Gaspard.  ``Comment on the paper ``Biology and nonequilibrium: remarks on a paper by J.L. England.'' by D. Ruelle.''  Eur. Phys. J. Special Topics, in reference [15].
\medskip\noindent
[9] C. Jarzynski.  ``Nonequilibrium equality for free energy differences.''  Phys. Rev. Lett. {\bf 78},2690-2693(1997)
\medskip\noindent
[10] J.Kurchan.  ``Fluctuation theorem for stochastic dynamics.''  J. Phys. A {\bf 31},3719-3729(1998).
\medskip\noindent
[11] R. Landauer.  ``Irreversibility and heat generation in the computing process.''  IBM J. Res. Dev. {\bf 5},183-191(1961).
\medskip\noindent
[12] J.L. Lebowitz and H. Spohn.  ``A Gallavotti-Cohen-type symmetry in the large deviation functional for stochastic dynamics.''  J. Statist. Phys. {\bf 95},333-365(1999).
\medskip\noindent
[13] C. Maes.  ``The fluctuation theorem as a Gibbs property''.  J. Stat. Phys. {\bf 95},367-392(1999)
\medskip\noindent
[14] D. Ruelle.  {\it Statistical Mechanics, Rigorous Results.}  Benjamin, New York, 1969.
\medskip\noindent
[15] D. Ruelle.  ``Smooth dynamics and new theoretical ideas in nonequilibrium statistical mechanics.''  J. Statist. Phys. {\bf 95},393-468(1999).
\medskip\noindent
[16] D. Ruelle.  ``Biology and nonequilibrium: remarks on a paper by J.L. England.''  Eur. Phys. J. Special Topics {\bf 224},935-945(2015).
\end